\documentclass[12pt,final,journal,letterpaper,oneside,onecolumn]{IEEEtran}
\usepackage[pdftex]{graphicx}
\pdfoutput=1
\usepackage[fleqn]{amsmath}

\def\f{\boldsymbol{f}}

\def\x{\boldsymbol{x}}
\def\y{\boldsymbol{y}}
\def\z{\boldsymbol{z}}
\def\A{\boldsymbol{A}}

\def\P{\boldsymbol{P}}

\def\X{\boldsymbol{X}}

\def\0{\boldsymbol{0}}
\def\cA{\mathcal A}

\def\cN{\mathcal N}
\def\cO{\mathcal O}

\def\cR{\mathcal R}

\def\cY{\mathcal Y}

\def\ov{\overline}

\def\wh{\widehat}
\def\wt{\widetilde}

\def\eqdef{{\displaystyle\mathop{=}^{\mbox{\rm def.}}}}

\newtheorem{theorem}{Theorem}

\begin{document}
\allowdisplaybreaks

\title{%
  A quick search method for audio signals
  based on a piecewise linear representation of feature trajectories
}%
\author{%
  Akisato~Kimura,~\IEEEmembership{Senior Member,~IEEE,}
  Kunio~Kashino,~\IEEEmembership{Senior Member,~IEEE,}
  Takayuki~Kurozumi,~\IEEEmembership{} and
  Hiroshi~Murase,~\IEEEmembership{Fellow,~IEEE}
  \thanks{%
    Manuscript received December 15, 2006; revised June 17, 2007;
    second revision September 24, 2007, Accepted October 6, 2007.
    The associate editor coordinating the review is Dr. Michael Goodwin.
  }%
  \thanks{%
    A. Kimura, K. Kashino and T. Kurozumi are with NTT Communication Science\
    Laboratories, NTT Corporation, Atsugi-shi 243-0198, Japan.
    E-mail: \{akisato, kunio, kurozumi\} $<$at$>$ eye brl ntt co jp
    URL: http://www.brl.ntt.co.jp/people/akisato/
  }%
  \thanks{%
    H. Murase is with the Graduate School of Information Science, Nagoya University,
    Nagoya-shi 464-8603, Japan.
    E-mail: murase $<$at$>$ is nagoya-u ac jp
  }%
  \thanks{%
    Some of the material in this paper was presented at the IEEE International
    Conference on Acoustics, Speech and Signal Processing (ICASSP2002), Orlando, FL, May
    2002, and at the IEEE International Conference on Multimedia and Expo (ICME2003),
    Baltimore, MD, June 2003.
  }%
}%

\maketitle

\begin{abstract}
This paper presents a new method for a quick similarity-based search through long
unlabeled audio streams to detect and locate audio clips provided by users. The method
involves feature-dimension reduction based on a piecewise linear representation of a
sequential feature trajectory extracted from a long audio stream. Two techniques enable
us to obtain a piecewise linear representation: the dynamic segmentation of feature
trajectories and the segment-based Karhunen-L\'{o}eve (KL) transform. The proposed search
method guarantees the same search results as the search method without the proposed
feature-dimension reduction method in principle. Experiment results indicate significant
improvements in search speed. For example the proposed method reduced the total search
time to approximately 1/12 that of previous methods and detected queries in
approximately 0.3 seconds from a 200-hour audio database.
\end{abstract}
\begin{IEEEkeywords}
audio retrieval, audio fingerprinting, content identification, feature trajectories,
piecewise linear representation, dynamic segmentation
\end{IEEEkeywords}

%

\section{Introduction}
\label{sec:intro}

This paper presents a method for searching quickly through unlabeled audio signal
archives (termed {\it stored signals}) to detect and locate given audio clips (termed
{\it query signals}) based on signal similarities.

\begin{figure}[t]
  \begin{center}
    \includegraphics[width=8cm]{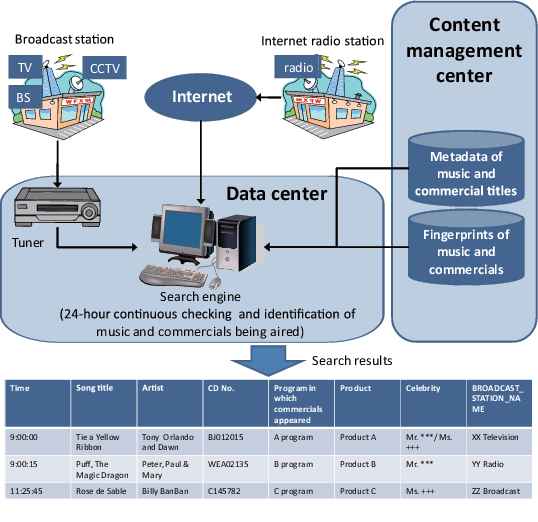}
  \end{center}
  \caption{
    Automatic monitoring system of broadcast content via music content identification.
  }
  \label{fig:meloQ}
\end{figure}

Many studies related to audio retrieval have dealt with content-based approaches
such as audio content classification \cite{AudioContentAnalysis,AudioClassify:Wold},
speech recognition \cite{AcousticIndex}, and music transcription
\cite{AcousticIndex,OverviewAudioRetrieval}. Therefore, these studies mainly focused on
associating audio signals with their meanings. In contrast, this study aims at achieving
a {\it similarity-based search} or more specifically {\it fingerprint identification},
which constitutes a search of and retrieval from unlabeled audio archives based only on a
signal similarity measure. That is, our objective is signal matching, not the association
of signals with their semantics. Although the range of applications for a
similarity-based search may seem narrow compared with content-based approaches, this is
not actually the case. The applications include the detection and statistical analysis of
broadcast music and commercial spots, and the content identification, detection and
copyright management of pirated copies of music clips. Fig. \ref{fig:meloQ} represents
one of the most representative examples of such applications, which has already been put
to practical use. This system automatically checks and identifies broadcast music clips
or commercial spots to provide copyright information or other detailed information about
the music or the spots.

In audio fingerprinting applications, the query and stored signals cannot be assumed to
be exactly the same even in the corresponding sections of the same sound, owing to, for
example, compression, transmission and irrelevant noises. Meanwhile, for the applications
to be practically viable, the features should be compact and the feature analysis should
be computationally efficient. For this purpose, several feature extraction methods have
been developed to attain the above objectives. 
Cano et al. \cite{RobustSoundModeling:Cano} modeled music segments as sequences of sound
classes estimated via unsupervised clustering and hidden Markov models (HMMs).
Burges et al. \cite{NoiseRobustFeatures:Burges} employed several layers of
Karhunen-L\'{o}eve (KL) transforms, which reduced the local statistical redundancy of
features with respect to time, and took account of robustness to shifting and pitching.
Oostveen et al. \cite{VideoIdentification:Phillips} represented each frame
of a video clip as a binary map and used the binary map sequence as a feature. This
feature is robust to global changes in luminance and contrast variations.
Haitsma et al. \cite{HighlyRobustAudioFingerprint} and Kurozumi et al.
\cite{ICASSP2007:kashino} each employed a similar approach in the context of audio
fingerprinting.
Wang \cite{IndustrialStrengthAudioSearch} developed a feature-point-based approach to
improve the robustness.
Our previous approach called the Time-series Active Search (TAS) method \cite{TASpaper}
introduced a histogram as a compact and noise-robust fingerprint, which
models the empirical distribution of feature vectors in a segment. Histograms are
sufficiently robust for monitoring broadcast music or detecting pirated copies. Another
novelty of this approach is its effectiveness in accelerating the search. Adjacent
histograms extracted from sliding audio segments are strongly correlated with each other.
Therefore, unnecessary matching calculations are avoided by exploiting the algebraic
properties of histograms.

Another important research issue regarding similarity-based approaches involves finding a
way to speed up
the search. Multi-dimensional indexing methods \cite{Rtree,SRtree} have frequently been
used for accelerating searches. However, when feature vectors are high-dimensional, as
they are typically with multimedia signals, the efficiency of the existing indexing
methods deteriorates significantly \cite{CostModelNN:berchtold,NNSearchAnalysis:weber}.
This is why search methods based on linear scans such as the TAS method are often
employed for searches with high-dimensional features. However, methods based solely on
linear scans may not be appropriate for managing large-scale signal archives, and
therefore dimension reduction should be introduced to mitigate this effect.

To this end, this paper presents a quick and accurate audio search method that uses
dimensionality reduction of histogram features. The method involves a piecewise linear
representation of histogram sequences by utilizing the continuity and local correlation
of the histogram sequences. A piecewise linear representation would be feasible for the
TAS framework since the histogram sequences form trajectories in multi-dimensional
spaces. By incorporating our method into the TAS framework, we significantly increase the
search speed while guaranteeing the same search results as the TAS method. We introduce
the following two techniques to obtain a piecewise representation: the dynamic
segmentation of the feature trajectories and the segment-based KL transform.

The segment-based KL transform involves the dimensionality reduction of divided
histogram sequences (called {\it segments}) by KL transform. We take advantage of the
continuity and local correlation of feature sequences extracted from audio signals.
Therefore, we expect to obtain a linear representation with few approximation errors
and low computational cost. The segment-based KL transform consists of the following
three components: The basic component of this technique reduces the dimensionality of
histogram features. The second component that utilizes residuals between original
histogram features and features after dimension reduction greatly reduces the required
number of histogram comparisons. Feature sampling is introduced as the third component.
This not only saves the storage space but also contributes to accelerating the search.

Dynamic segmentation refers to the division of histogram sequences into segments of
various lengths to achieve the greatest possible reduction in the average dimensionality
of the histogram features. One of the biggest problems in dynamic segmentation is that
finding the optimal set of partitions that minimizes the average dimensionality requires
a substantial calculation. The computational time must be no more than that
needed for capturing audio signals from the viewpoint of practical applicability. To
reduce the calculation cost, our technique addresses the quick suboptimal partitioning
of the histogram trajectories, which consists of local optimization to avoid recursive
calculations and the coarse-to-fine detection of segment boundaries.

This paper is organized as follows: Section \ref{sec:pre} introduces the notations
and definitions necessary for the subsequent explanations. Section \ref{sec:TAS} explains
the TAS method upon which our method is founded. Section \ref{sec:proposed} outlines
the proposed search method. Section \ref{sec:piecewise} discusses a
dimensionality reduction technique with the segment-based KL transform. Section
\ref{sec:dynamic} details dynamic segmentation. Section \ref{sec:exp} presents
experimental results related to the search speed and shows the advantages of the proposed
method. Section \ref{sec:discuss} further discusses the advantages and shortcomings of
the proposed method as well as providing additional experimental results. Section
\ref{sec:conclusion} concludes the paper.

\section{Preliminaries}\label{sec:pre}

Let $\cN$ be the set of all non-negative numbers, $\cR$ be the set of all real numbers,
and $\cN^n$ be a $n$-ary Cartesian product of $\cN$.
Vectors are denoted by boldface lower-case letters, e.g. $\x$, and matrices are denoted
by
boldface upper-case letters, e.g. $\A$. The superscript $t$ stands for the transposition
of a vector or a matrix, e.g. $\x^t$ or $\A^t$. The Euclidean norm of an $n$-dimensional
vector $\x\in\cR^n$ is denoted as $\|\x\|$:
\begin{eqnarray*}
  \|\x\| &\eqdef& \left(\sum_{i=1}^n |x_i|^2\right)^{1/2},
\end{eqnarray*}
where $|x|$ is the magnitude of $x$.
For any function $f(\cdot)$ and a random variable $X$, $E[f(X)]$ stands for the
expectation of $f(X)$. Similarly, for a given value $y\in\cY$, some function
$g(\cdot,\cdot)$ and a random variable $X$, $E[f(X,y)|y]$ stands for the conditional
expectation of $g(X,y)$ given $y$.

\section{Time-series Active Search}\label{sec:TAS}


\begin{figure}[t]
  \begin{center}
    \includegraphics[width=8cm]{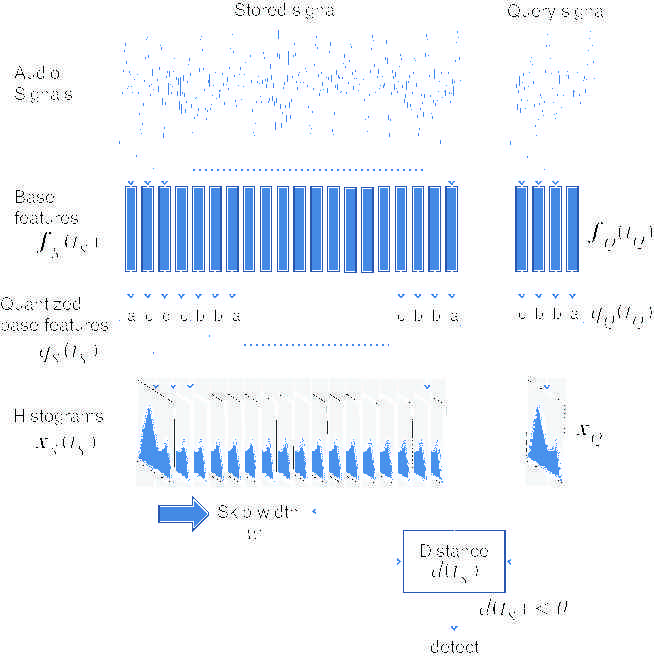}
  \end{center}
  \caption{Overview of the Time-series Active Search (TAS) method.}
  \label{fig:TAS}
\end{figure}

Fig. \ref{fig:TAS} outlines the Time-series Active Search (TAS) method, which is the
basis of our proposed method. We provide a summary of the algorithm here. Details can
be found in \cite{TASpaper}.

\noindent
[Preparation stage]
\begin{enumerate}
\item
  Base features are extracted from the stored signal. Our preliminary experiments showed
  that the short-time frequency spectrum provides sufficient accuracy for our
  similarity-based search task. Base features are extracted at every sampled time step,
  for example, every 10 msec. Henceforth, we call the sampled points {\it frames} (the
  term was inspired by video frames). Base features are denoted as $\f_S(t_S)$
  $(0\le t_S<L_S)$, where $t_S$ represents the position in the stored signal and $L_S$ is
  the length of the stored signal (i.e. the number of frames in the stored signal).
\item
  Every base feature is quantized by vector quantization (VQ). A codebook
  $\{\ov{\f}_i\}_{i=1}^{n}$ is created beforehand, where $n$ is the codebook
  size (i.e. the number of codewords in the codebook). We utilize the Linde-Buzo-Gray
  (LBG) algorithm \cite{VQ:gray} for codebook creation. A quantized base feature
  $q_S(t_S)$ is expressed as a VQ codeword assigned to the corresponding base feature
  $\f_S(t_S)$, which is determined as
  \begin{eqnarray*}
    q_S(t_S) &=& \arg\min_{1\le i\le n}\|\f_S(t_S)-\ov{\f}_i\|^2.
  \end{eqnarray*}
\end{enumerate}

\noindent
[Search stage]
\begin{enumerate}
\item
  Base features $\f_Q(t_Q)$ $(0\le t_Q< L_Q)$ of the query signal are extracted in the
  same way as the stored signal and quantized with the codebook $\{\ov{\f}_i\}_{i=1}^{n}$
  created in the preparation stage, where $t_Q$ represents the position in the query
  signal and $L_Q$ is its length. We do not have to take into account the calculation
  time for feature quantization since it takes less than 1\% of the length of the signal.
  A quantized base feature for the query signal is denoted as $q_Q(t_Q)$.
\item
  Histograms are created; one for the stored signal denoted as $\x_S(t_S)$ and the other
  for the query signal denoted as $\x_Q$ . First, windows are applied to the sequences of
  quantized base features extracted from the query and stored signals. The window length
  $W$ (i.e. the number of frames in the window) is set at $W=L_Q$, namely the length of
  the query signal. A histogram is created by counting the instances of each VQ codeword
  over the window. Therefore, each index of a histogram bin corresponds to a VQ codeword.
  We note that a histogram does not take the codeword order into account.
\item
  Histogram matching is executed based on the distance between histograms, computed as
  \begin{eqnarray*}
    d(t_S) &\eqdef& \|\x_S(t_S)-\x_Q\|.
  \end{eqnarray*}
  When the distance $d(t_S)$ falls below a given value ({\it search threshold})
  $\theta$, the query signal is considered to be detected at the position $t_S$ of the
  stored signal.
\item
  A window on the stored signal is shifted forward in time and the procedure returns to
  Step 2). As the window for the stored signal shifts forward in time, VQ codewords
  included in the window cannot change so rapidly, which means that histograms cannot
  also change so rapidly. This implies that for a given positive integer $w$ the
  lower bound on the distance $d(t_S+w)$ is obtained from the triangular inequality as
  follows:
  \begin{eqnarray*}
    d(t_S+w) &\ge& \max\{0,d(t_S)-\sqrt{2}w\},
  \end{eqnarray*}
  where $\sqrt{2}$ is the maximum distance between $\x_S(t_S)$ and $\x_S(t_S+w)$.
  Therefore, the skip width $w(t_S)$ of the window at the $t_S$-th frame is obtained as
  \begin{eqnarray}
    \lefteqn{w(t_S)}\nonumber\\
    &=& \left\{
    \begin{array}{ll}
      \mbox{floor}\left({\displaystyle\frac{d(t_S)-\theta}{\sqrt{2}}}\right)+1
      & (\mbox{if }d(t_S)>\theta) \\
      1, & (\mbox{otherwise})
    \end{array}
    \right.
    \label{eq:L2skip}
  \end{eqnarray}
  where $\mbox{floor}(a)$ indicates the largest integer less than $a$. We note that no
  sections will ever be missed that have distance values smaller than the search
  threshold $\theta$, even if we skip the width $w(t_S)$ given by Eq. (\ref{eq:L2skip}). 
\end{enumerate}

\section{Framework of proposed search method}
\label{sec:proposed}

\begin{figure}[t]
  \begin{center}
    \includegraphics[width=8cm]{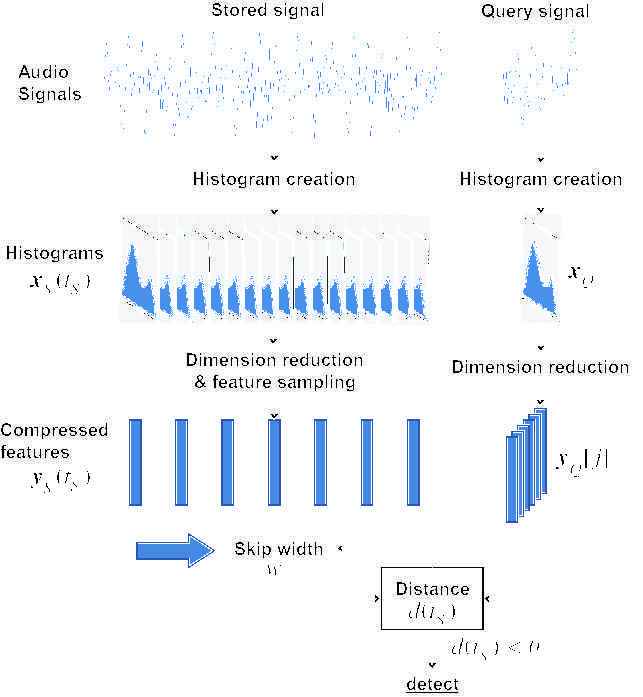}
  \end{center}
  \caption{Overview of proposed search method.}
  \label{fig:outline:proposed}
\end{figure}
\begin{figure}[t]
  \begin{center}
    \includegraphics[width=8cm]{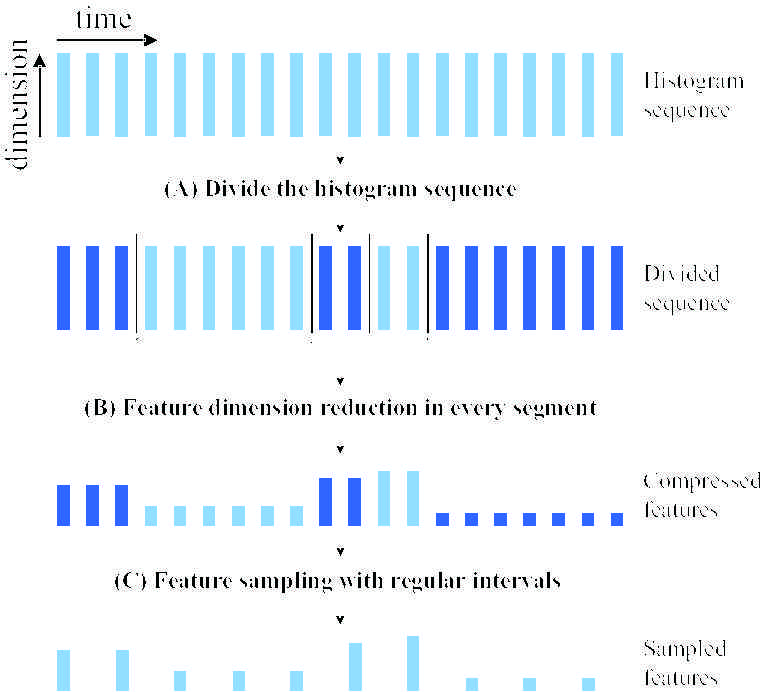}
  \end{center}
  \caption{Overview of the procedure for obtaining compressed features.}
  \label{fig:compress_scheme}
\end{figure}

The proposed method improves the TAS method so that the search is accelerated without
{\it false dismissals} (incorrectly missing segments that should be detected) or
{\it false detections} (identifying incorrect matches). To accomplish this, we introduce
feature-dimension reduction as explained in Sections \ref{sec:piecewise} and
\ref{sec:dynamic}, which reduces the calculation costs required for matching.

Fig. \ref{fig:outline:proposed} shows an overview of the proposed search method, and
Fig. \ref{fig:compress_scheme} outlines the procedure for feature-dimension reduction.
The procedure consists of a preparation stage and a search stage.

\noindent
[Preparation stage]
\begin{enumerate}
  \item Base features $\f_S(t_S)$ are extracted from the stored signal and quantized, to
        create quantized base features $q_S(t_S)$. The procedure is the same as that of
	the TAS method.
  \item Histograms $\x_S(t_S)$ are created in advance from the quantized base features of
        the stored signal by shifting
	a window of a predefined length $W$. We note that with the TAS method the
	window length $W$ varies from one search to another, while with the present
	method the window length $W$ is fixed. This is because histograms $\x_S(t_S)$ for
	the stored signal are created prior to the search. We should also note that the
	TAS method does not create histograms prior to the search because sequences of VQ
	codewords need much less storage space than histogram sequences.
  \item A piecewise linear representation of the extracted histogram sequence is
        obtained (Fig. \ref{fig:compress_scheme} block (A)). This representation is
	characterized by a set $T=\{t_j\}_{j=0}^M$ of segment boundaries expressed by
	their frame numbers and a set $\{p_j(\cdot)\}_{j=1}^M$ of $M$ functions, where
	$M$ is the number of segments, $t_0=0$ and $t_M=L_S$. The $j$-th segment is
	expressed as a half-open interval $[t_{j-1},t_j)$ since it starts from
	$\x_S(t_{j-1})$ and ends at $\x_S(t_j-1)$. Section \ref{sec:dynamic} shows
	how to obtain such segment boundaries. Each function $p_j(\cdot):$
	$\cN^n\to\cR^{m_j}$ that corresponds to the $j$-th segment reduces the
	dimensionality $n$ of the histogram to the dimensionality $m_j$. Section
	\ref{subsec:piecewise:piecewise} shows how to determine these functions.
  \item The histograms $\x_S(t_S)$ are compressed by using the functions
        $\{p_j(\cdot)\}_{j=1}^M$ obtained in the previous step, and then {\it compressed
        features} $\y_S(t_S)$ are created (Fig. \ref{fig:compress_scheme} block (B)).
        Section \ref{subsec:piecewise:bound} details how to create compressed features.
  \item The compressed features $\y_S(t_S)$ are sampled at regular intervals
        (Fig. \ref{fig:compress_scheme} block (C)). The details are presented in Section
        \ref{subsec:piecewise:sampling}.
\end{enumerate}

\noindent
[Search stage]
\begin{enumerate}
  \item Base features $\f_Q(t_Q)$ are extracted and a histogram $\x_Q$ is created from
        the query signal in the same way as the TAS method.
  \item The histogram $\x_Q$ is compressed based on the functions
        $\{p_j(\cdot)\}_{j=1}^M$ obtained in the preparation stage, to create $M$
        compressed features $\y_Q[j]$ $(j=1,\cdots,M)$. Each compressed feature $\y_Q[j]$
	corresponds to the $j$-th function $p_j(\cdot)$. The procedure used to create
	compressed features is the same as that for the stored signal.
  \item Compressed features created from the stored and query signals are matched,
        that is, the distance $\wt{d}(t_S)=\|\y_S(t_S)-\y_Q[j_{t_S}]\|$ between two
	compressed
	features $\y_S(t_S)$ and $\y_Q[j_{t_S}]$ is calculated, where $j_{t_S}$
	represents the index of the segment that contains $\x_S(t_S)$, namely
	$t_{j_{t_S}-1}\le t_S< t_{j_{t_S}}$.
  \item If the distance falls below the search threshold $\theta$, the original
        histograms $\x_S(t_S)$ corresponding to the surviving compressed features
	$\y_S(t_S)$ are verified. Namely, the distance $d(t_S)=\|\x_S(t_S)-\x_Q\|$ is
	calculated and compared with the search threshold $\theta$.
  \item A window on the stored signal is shifted forward in time and the procedure goes
        back to Step 3). The skip width of the window is calculated from the distance
	$\wt{d}(t_S)$ between compressed features.
\end{enumerate}

\section{Dimension reduction based on piecewise linear representation}
\label{sec:piecewise}
\subsection{Related work}
\label{subsec:piecewise:related}

In most practical similarity-based searches, we cannot expect the features to be globally
correlated, and therefore there is little hope of reducing dimensionality over entire
feature spaces. However, even when there is no global correlation, feature subsets may
exist that are locally correlated. Such local correlation of feature subsets has the
potential to further reduce feature dimensionality.

A large number of dimensionality reduction methods have been proposed that focused on
local correlation (e.g. \cite{VQPCA,mixPCA:bishop,LLE,PROCLUS}). Many of these methods do
not assume any specific characteristics. Now, we are concentrating on the dimensionality
reduction of time-series signals, and therefore we take advantage of their continuity and
local correlation. The computational cost for obtaining such feature subsets is expected
to be very small compared with that of existing methods that do not utilize the
continuity and local correlation of time-series signals.

Dimensionality reduction methods for time-series signals are categorized into two types:
{\it temporal dimensionality reduction}, namely dimensionality reduction along the
temporal axis (e.g. feature sampling), and {\it spatial dimensionality reduction}, namely
the dimensionality
reduction of each multi-dimensional feature sample. Keogh {\it et al.} \cite{APCA,PAA}
and Wang {\it et al.} \cite{TimeSeriesApprox:wang} have introduced temporal
dimensionality reduction into waveform signal retrieval. Their framework considers the
waveform itself as a feature for detecting similar signal segments. That is why they
mainly focused on temporal dimensionality reduction. When considering audio
fingerprinting, however, we handle sequences of high-dimensional features that are
necessary to
identify various kinds of audio segments. Thus, both spatial and temporal dimensionality
reduction are required. To this end, our method mainly focuses on spatial
dimensionality reduction. We also incorporate a temporal dimensionality reduction
technique inspired by the method of Keogh et al. \cite{PAA}, which is described in
Section \ref{subsec:piecewise:sampling}.

\subsection{Segment-based KL transform}
\label{subsec:piecewise:piecewise}

\begin{figure}[t]
  \begin{center}
    \includegraphics[width=8cm]{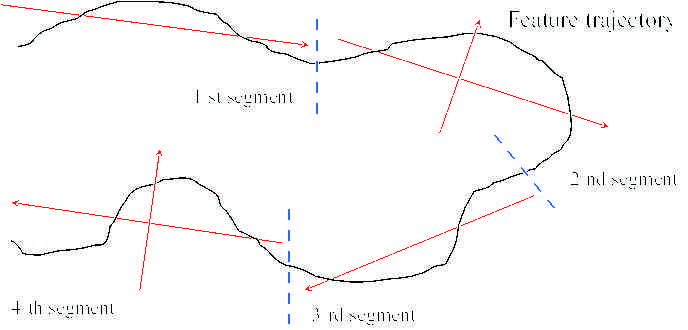}
  \end{center}
  \caption{Intuitive illustration of piecewise linear representation.}
  \label{fig:piecewise}
\end{figure}

Fig. \ref{fig:piecewise} shows an intuitive example of a piecewise linear representation.
Since the histograms are created by shifting the window forward in time, successive
histograms cannot change rapidly. Therefore, the histogram sequence forms a smooth
trajectory in an $n$-dimensional space even if a stored audio signal includes distinct
non-sequential patterns, such as irregular drum beats and intervals between music clips.
This implies that a piecewise lower-dimensional representation is feasible for such a
sequential histogram trajectory.

As the first step towards obtaining a piecewise representation, the histogram sequence is
divided into $M$ segments. Dynamic segmentation is introduced here, which enhances
feature-dimension reduction performance. This will be explained in detail in Section
\ref{sec:dynamic}. Second, a KL transform is performed for every segment and a minimum
number of eigenvectors are selected such that the sum of their {\it contribution rates}
exceeds a predefined value $\sigma$, where the contribution rate of an eigenvector
stands for its eigenvalue divided by the sum of all eigenvalues, and the predefined
value $\sigma$ is called the {\it contribution threshold}.
The number of selected eigenvectors in the $j$-th segment is written as
$m_j$. Then, a function $p_j(\cdot):$ $\cN^n\to\cR^{m_j}$ $(j=1,2,\cdots,M)$ for
dimensionality reduction is determined as a map to a subspace whose bases are the
selected eigenvectors:
\begin{eqnarray}
  p_j(\x) &=& \P_j^t(\x-\overline{\x}_j),
\end{eqnarray}
where $\x$ is a histogram, $\overline{\x}_j$ is the centroid of histograms contained in
the $j$-th segment, and $\P_j$ is an ($n\times m_j$) matrix whose columns are the
selected eigenvectors. Finally, each histogram is compressed by using the function
$p_j(\cdot)$ of the segment to which the histogram belongs. Henceforth, we refer to
$p_j(\x)$ as a {\it projected feature} of a histogram $\x$.

In the following, we omit the index $j$ corresponding to a segment unless it is
specifically needed, e.g. $p(\x)$ and $\ov{\x}$.

\subsection{Distance bounding}
\label{subsec:piecewise:bound}

From the nature of the KL transform, the distance between two projected features gives
the
lower bound of the distance between corresponding original histograms. However, this
bound does not approximate the original distance well, and this results in many false
detections.

To improve the distance bound, we introduce a new technique. Let us define a
{\it projection
distance} $\delta(p,\x)$ as the distance between a histogram $\x$ and the corresponding
projected feature $\z=p(\x)$:
\begin{eqnarray}
  \delta(p,\x) &\eqdef& \|\x-q(\z)\|,
    \label{eq:projdist}
\end{eqnarray}
where $q(\cdot): \cR^m\to\cR^n$ is the generalized inverse map of $p(\cdot)$, defined
as
\begin{eqnarray*}
  q(\z) &\eqdef& \P\z+\overline{\x}.
\end{eqnarray*}
Here we create a compressed feature $\y$, which is the projected feature
$\z=(z_1,z_2,\cdots,z_m)^t$ along with the projection distance $\delta(p,\x)$:
\begin{eqnarray*}
  \y = \y(p,\x) &=& (z_1,z_2,\cdots,z_m,\delta(p,\x))^t,
\end{eqnarray*}
where $\y(p,\x)$ means that $\y$ is determined by $p$ and $\x$. The Euclidean distance
between compressed features is utilized as a new criterion for matching instead of the
Euclidean distance between projected features. The distance is expressed as
\begin{eqnarray}
  \lefteqn{\|\y_S-\y_Q\|^2}\nonumber\\
  &=& \|\z_S-\z_Q\|^2+\{\delta(p,\x_S)-\delta(p,\x_Q)\}^2,
  \label{eq:distlb}
\end{eqnarray}
where $\z_S=p(\x_S)$ (resp. $\z_Q=p(\x_Q)$) is the project feature derived from the
original histograms $\x_S$ (resp. $\x_Q$) and $\y_S=\y_S(p,\x_S)$ (resp.
$\y_Q=\y_Q(p,\x_Q)$) is the corresponding compressed feature. Eq. (\ref{eq:distlb})
implies that the distance between compressed features is larger than the distance between
corresponding projected features. In addition, from the above discussions, we have the
following two properties, which indicate that the distance $\|\y_S-\y_Q\|$ between two
compressed features is a better approximation of the distance $\|\x_S-\x_Q\|$ between the
original histograms than the distance $\|\z_S-\z_Q\|$ between projected features (Theorem
\ref{theorem:bound}), and the expected approximation error is much smaller (Theorem
\ref{theorem:error}).

\begin{figure}[t]
  \begin{center}
    \includegraphics[width=8cm]{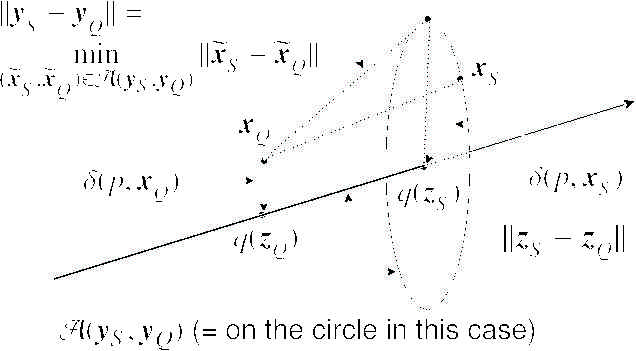}
  \end{center}
  \caption{
    Intuitive illustration of relationships between projection distance, distance between
    projected features and distance between compressed features.
  }
  \label{fig:project}
\end{figure}

\begin{theorem} \label{theorem:bound}
  \begin{eqnarray}
    \lefteqn{\|\z_S-\z_Q\|\le \|\y_S-\y_Q\|}\nonumber\\
    &=& \min_{(\tilde{\x}_S,\tilde{\x}_Q)\in\cA(\y_S,\y_Q)}\|\tilde{\x}_S-\tilde{\x}_Q\|
        \le \|\x_S-\x_Q\|, \label{eq:ineq2}
  \end{eqnarray}
  where $\cA(\y_S,\y_Q)$ is the set of all possible pairs $(\tilde{\x}_S,\tilde{\x}_Q)$
  of original histograms for given compressed features $(\y_S,\y_Q)$.
\end{theorem}

\begin{theorem} \label{theorem:error}
  Suppose that random variables $(X_S^n,X_Q^n)$ corresponding to the original histograms
  $(\x_S,\x_Q)$ have a uniform distribution on the set $\cA(\y_S,\y_Q)$
  defined in Theorem \ref{theorem:bound}, and $E[\delta(p,X_S^n)]\gg E[\delta(p,X_Q^n)]$.
  The expected approximation errors can be evaluated as
  \begin{eqnarray}
    \lefteqn{E\left[\left.\|X_S^n-X_Q^n\|^2-\|\y_S-\y_Q\|^2\right|
	\y_S,\y_Q\right]}\nonumber\\
    &\ll& E\left[\left.\|X_S^n-X_Q^n\|^2-\|\z_S-\z_Q\|^2\right|\y_S,\y_Q\right].
          \label{eq:error2}
  \end{eqnarray}
\end{theorem}

\medskip
The proofs are shown in the appendix. Fig. \ref{fig:project} shows an intuitive
illustration of the relationships between projection distances, distances between
projected features and distances between compressed features, where the histograms are in
a 3-dimensional space and the subspace dimensionality is 1. In this case, for given
compressed features $(\y_S,\y_Q)$ and a fixed query histogram $\x_Q$, a stored histogram
$\x_S$ must be on a circle whose center is $q(\z_Q)$. This circle corresponds to the
set $\cA(\y_S,\y_Q)$.

\subsection{Feature sampling}
\label{subsec:piecewise:sampling}

In the TAS method, quantized base features are stored, because they need much less
storage space than the histogram sequence and creating histograms on the spot takes
little calculation. With the present method, however, compressed features must be
computed and stored in advance so that the search results can be returned as quickly as
possible, and therefore much more storage space is needed than with
the TAS method. The increase in storage space may cause a reduction in search speed due
to the increase in disk access.

Based on the above discussion, we incorporate feature sampling in the temporal domain.
The following idea is inspired by the technique called Piecewise Aggregate Approximation
(PAA) \cite{PAA}. With the proposed feature sampling method, first a compressed feature
sequence $\{\y_S(t_S)\}_{t_S=0}^{L_S-W-1}$ is divided into subsequences 
\begin{eqnarray*}
  \{\y_S(ia),\y_S(ia+1),\cdots,\y_S(ia+a-1)\}_{i=0,1,\cdots}
\end{eqnarray*}
of length $a$. Then, the first compressed feature $\y_S(ia)$ of every subsequence is
selected as a {\it representative feature}. A lower bound of the distances between the
query and stored compressed features contained in the subsequence can be expressed in
terms of the representative feature $\y_S(ia)$. This bound is obtained from the
triangular inequality as follows:
\begin{eqnarray*}
  && \|\y_S(ia+k)-\y_Q\| \ge \|\y_S(ia)-\y_Q\|-\ov{d}(i),\\
  && \ov{d}(i) \eqdef \max_{0\le k'\le a-1}\|\y_S(ia+k')-\y_S(ia)\|.\\
  && \hspace{15mm}(\forall i=0,1,\cdots,\quad\forall k=0,\cdots,a-1)
\end{eqnarray*}
This implies that preserving the representative feature $\y_S(ia)$ and the maximum
distance $\ov{d}(i)$ is sufficient to guarantee that there are no false dismissals. 

This feature sampling is feasible for histogram sequences because successive histograms
cannot change rapidly. Furthermore, the technique mentioned in this section will also
contribute to accelerating the search, especially when successive histograms change
little.

\section{Dynamic segmentation}
\label{sec:dynamic}
\subsection{Related work}
\label{subsec:dynamic:related}

The approach used for dividing histogram sequences into segments is critical for
realizing efficient feature-dimension reduction since the KL transform is most effective
when the constituent elements in the histogram segments are similar. To achieve this, we
introduce a dynamic segmentation strategy.

Dynamic segmentation is a generic term that refers to techniques for dividing sequences
into segments of various lengths. Dynamic segmentation methods for time-series signals
have already been applied to various kinds of applications such as speech coding (e.g.
\cite{SegmentVQBasedLPCCoding}), the temporal compression of waveform signals
\cite{AcousticPatternSegmentation}, the automatic segmentation of speech signals into
phonic units \cite{DynamicSegmentation:Svendsen}, sinusoidal modeling of audio signals
\cite{PhD:goodwin,SinusoidalModeling:goodwin,SinusoidalModeling:jang} and motion
segmentation in video signals \cite{DynamicSegmentation:Mann}. We employ dynamic
segmentation to minimize the average dimensionality of high-dimensional feature
trajectories.

Dynamic segmentation can improve dimension reduction performance. However, finding the
optimal boundaries still requires a substantial calculation. With this in mind, several
studies have adopted suboptimal approaches, such as longest line fitting
\cite{TimeSeriesApprox:wang}, wavelet decomposition \cite{TimeSeriesApprox:wang,APCA} and
the bottom-up merging of segments \cite{ProbabilisticMatching}. The first two approaches
still incur a substantial calculation cost for long time-series signals. The last
approach is promising as regards obtaining a rough global approximation at a practical
calculation cost. This method is compatible with ours, however, we mainly focus on a more
precise local optimization.

\subsection{Framework}
\label{subsec:dynamic:framework}

\begin{figure}[t]
  \begin{center}
    \includegraphics[width=8cm]{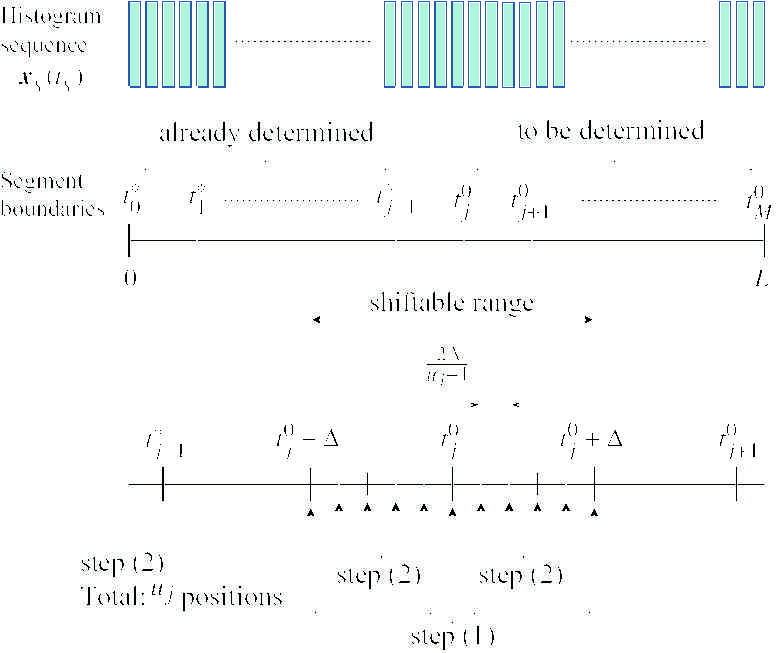}
    \caption{Outline of dynamic segmentation.}
    \label{fig:segment}
  \end{center}
\end{figure}

Fig. \ref{fig:segment} shows an outline of our dynamic segmentation method. The
objective of the dynamic segmentation method is to divide the stored histogram sequence
so that its piecewise linear representation is well characterized by a set of lower
dimensional subspaces. To this end, we formulate the dynamic segmentation as a way to
find a set $T^*=\{t_j^*\}_{j=0}^M$ of segment boundaries that minimize the average
dimensionality of these segment-approximating subspaces on condition that the boundary
$t_j^*$ between the $j$-th and the $(j+1)$-th segments is in a {\it shiftable range}
$S_j$, which is defined as a section with a width $\Delta$ in the vicinity of the initial
position $t_j^0$ of the boundary between the $j$-th and the $(j+1)$-th segments. Namely,
the set $T^*$ of the optimal segment boundaries is given by the following formula:
\begin{eqnarray}
  T^*
  &   =  & \left\{t_j^*\right\}_{j=0}^M \nonumber\\
  &\eqdef& \hspace{-4mm}
           {\displaystyle\mathop{\arg\min}_{\{t_j\}_{j=0}^M:t_j\in S_j\forall j}}
	   \frac{1}{L_S}\sum_{i=1}^M(t_j-t_{j-1})\cdot c(t_{j-1},t_j,\sigma)
           \label{eq:defdynamic}\\
  S_j &{\displaystyle\mathop=^{\rm def.}}& \{t_j: t_j^0-\Delta\le t_j\le t_j^0+\Delta\}
      \label{eq:shiftable}
\end{eqnarray}
where $c(t_i,t_j,\sigma)$ represents the subspace dimensionality on the segment between
the $t_i$-th and the $t_j$-th frames for a given contribution threshold $\sigma$, 
$t_0^*=0$ and $t_M^*=L_S$. The initial positions of the segment boundaries are set
beforehand by equi-partitioning.

The above optimization problem defined by Eq. (\ref{eq:defdynamic})
would normally be solved with dynamic programming (DP) (e.g.
\cite{DynamicProgramming}). However, DP is not practical in this case. Deriving
$c(t_{j-1},t_j,\sigma)$ included in Eq. (\ref{eq:defdynamic}) incurs a substantial
calculation cost since it is equivalent to executing a KL transform calculation for the
segment $[t_{j-1},t_j)$. This implies that the DP-based approach requires a significant
amount of calculation, although less than a naive approach. The above discussion implies
that we should reduce the number of KL transform calculations to reduce the total
calculation cost required for the optimization. When we adopt the total number of KL
transform calculations as a measure for assessing the calculation cost, the cost is
evaluated as $\cO(M\Delta^2)$, where $M$ is the number of segments and $\Delta$ is the
width of the shiftable range.

To reduce the calculation cost, we instead adopt a suboptimal approach. Two techniques
are incorporated: local optimization and the coarse-to-fine detection of segment
boundaries. We explain these two techniques in the following sections.

\subsection{Local optimization}
\label{subsec:dynamic:local}

The local optimization technique modifies the formulation (Eq. (\ref{eq:defdynamic})) of
dynamic segmentation so that it minimizes the average dimensionality of the subspaces
of adjoining segments. The basic idea is similar to the ``forward segmentation''
technique introduced by Goodwin \cite{PhD:goodwin,SinusoidalModeling:goodwin} for
deriving accurate sinusoidal models of audio signals. The position $t_j^*$ of the
boundary is determined by using the following forward recursion as a substitute for Eq.
(\ref{eq:defdynamic}):
\begin{eqnarray}
  t_j^*
  &=& \arg\min_{t_j\in S_j}
      \frac{(t_j-t_{j-1}^*)c_j^*+(t_{j+1}^0-t_j)c_{j+1}^0}{t_{j+1}^0-t_{j-1}^*},
      \label{eq:formulation}
\end{eqnarray}
which is here given by
\begin{eqnarray*}
  && c_j^*=c(t_{j-1}^*,t_j,\sigma),\ c_{j+1}^0=c(t_j,t_{j+1}^0,\sigma),
\end{eqnarray*}
and $S_j$ is defined in Eq. (\ref{eq:shiftable}). As can be seen in Eq.
(\ref{eq:formulation}), we can determine each segment boundary independently, unlike the
formulation of Eq. (\ref{eq:defdynamic}). Therefore, the local optimization
technique can reduce the amount of calculation needed for extracting an appropriate
representation, which is evaluated as $\cO(M\Delta)$, where $M$ is the number of segments
and $\Delta$ is the width of the shiftable range.

\subsection{Coarse-to-fine detection}\label{subsec:dynamic:coarse}

\begin{figure}[t]
  \begin{center}
    \includegraphics[width=8cm]{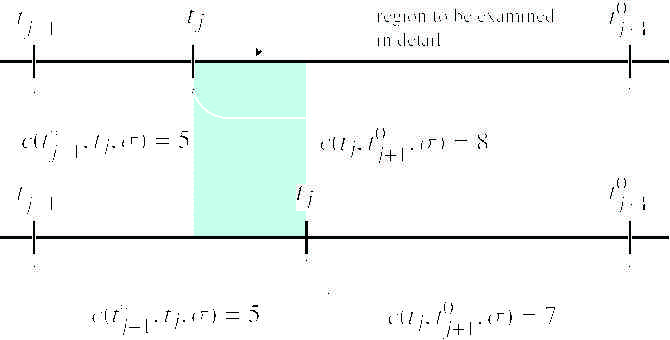}
    \caption{
      Example 1: $c(t_j,t_{j+1}^0,\sigma)$ decreases when the boundary $t_j$ is shifted
      forward in time.
    }\label{fig:coarse1}~\\
    \medskip
    \includegraphics[width=8cm]{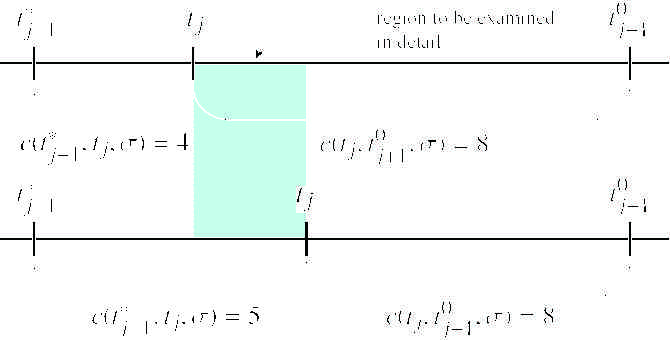}
    \caption{
      Example 2: $c(t_{j-1}^*,t_j,\sigma)$ increases when the boundary $t_j$ is shifted
      forward in time.
    }\label{fig:coarse2}
  \end{center}
\end{figure}

The coarse-to-fine detection technique selects suboptimal boundaries in the sense of Eq.
(\ref{eq:formulation}) with less computational cost. We note that small boundary shifts
do not contribute greatly to changes in segment dimensionality because
successive histograms cannot change rapidly. With this in mind, we assume that the
optimal positions of the segment boundaries are at the edges of the shiftable range or at
the points where dimensions change. Figs. \ref{fig:coarse1} and \ref{fig:coarse2} show
two intuitive examples where the optimal position of the segment boundary may be at
the point where dimensionality changes.
The coarse-to-fine detection technique quickly finds the points where the dimensions
change. The procedure for this technique has three steps.
\begin{enumerate}
\item
  The dimensions of the $j$-th and $(j+1)$-th segments are calculated when the segment
  boundary $t_j$ is at the initial position $t_j^0$ and the edges ($t_j^0-\Delta$ and
  $t_j^0+\Delta$) of its shiftable range.
\item
  The dimensions of the $j$-th and $(j+1)$-th segments are calculated when the segment
  boundary $t_j$ is at the position $t_j^0-\Delta+\frac{2\Delta}{u_j+1}i$
  $(i=1,2,\cdots,u_j)$, where $u_j$ determines the number of calculations in this step.
\item
  The dimensions of the $j$-th and $(j+1)$-th segments are calculated in detail when the
  segment boundary $t_j$ is in the positions where dimension changes are detected in
  the previous step.
\end{enumerate}
We determine the number $u_j$ of dimension calculations in step 2 so that the number of
calculations in all the above steps, $f_j(u_j)$, is minimized. Then, $f_j(u_j)$ is given
as follows:
\begin{eqnarray*}
  f_j(u_j) = 2\left((3+u_j)+K_j\frac{\Delta}{\frac 12 u_j+1}\right),
\end{eqnarray*}
where $K_j$ is the estimated number of positions where the dimensionalities change, which
is experimentally determined as
\begin{eqnarray*}
  K_j &=& c_{LR}-c_{LL},\\
      & & (\mbox{if } c_{LR}\le c_{RR}, c_{LL}<c_{RL})\\
  K_j &=& (c_{LC}-c_{LL})+\min(c_{RC},c_{LR})-\min(c_{LC},c_{RR}),\\
      & & (\mbox{if } c_{LR}>c_{RR},c_{LL}<c_{RL},c_{LC}\le c_{RC})\\
  K_j &=& (c_{RC}-c_{RR})+\min(c_{LC},c_{RL})-\min(c_{RC},c_{LL}),\\
      & & (\mbox{if } c_{LR}>c_{RR},c_{LL}<c_{RL},c_{LC}>c_{RC})\\
  K_j &=& c_{RL}-c_{RR},  \quad(\mbox{Otherwise})
\end{eqnarray*}
and
\begin{eqnarray*}
  \begin{array}{ll}
  c_{LL}=c(t_{j-1}^*,t_j^0-\Delta,\sigma),& c_{RL}=c(t_j^0-\Delta,t_{j+1}^0,\sigma),\\
  c_{LC}=c(t_{j-1}^*,t_j^0,       \sigma),& c_{RC}=c(t_j^0       ,t_{j+1}^0,\sigma),\\
  c_{LR}=c(t_{j-1}^*,t_j^0+\Delta,\sigma),& c_{RR}=c(t_j^0+\Delta,t_{j+1}^0,\sigma).
  \end{array}
\end{eqnarray*}
The first term of $f_j(u_j)$ refers to the number of calculations in steps 1 and 2, and
the second term corresponds to that in step 3. $f_j(u_j)$ takes the minimum value
$4\sqrt{2K_j\Delta}+2$ when $u_j=\sqrt{2K_j\Delta}-2$.
The calculation cost when incorporating local optimization and coarse-to-fine detection
techniques is evaluated as follows:
\begin{eqnarray*}
  E\left[M\left(4\sqrt{2K_j\Delta}+2\right)\right]
  &\le& M\left(4\sqrt{2K\Delta}+2\right) \\
  & = & {\cal O}\left(M\sqrt{K\Delta}\right),
\end{eqnarray*}
where $K=E[K_j]$, $M$ is the number of segments and $\Delta$ is the width of the
shiftable range. The first inequality is derived from Jensen's inequality (e.g.
\cite[Theorem 2.6.2]{CoverThomas}). The coarse-to-fine detection technique can
additionally reduce the calculation cost because $K$ is usually much smaller than
$\Delta$.

\section{Experiments}
\label{sec:exp}
\subsection{Conditions}
\label{sec:exp:conditions}

We tested the proposed method in terms of calculation cost in relation to search speed.
We again note that the proposed search method guarantees the same search results as the
TAS method in principle, and therefore we need to evaluate the search speed. The
search accuracy for the TAS method was reported in a previous paper \cite{TASpaper}. In
summary, for audio identification tasks, there were no false detections or false
dismissals down to an S/N ratio of 20 dB if the query duration was longer than 10
seconds.

In the experiments, we used a recording of a real TV broadcast. An audio signal
broadcast from a particular TV station was  recorded and encoded in MPEG-1
Layer 3 (MP3) format. We recorded a 200-hour audio signal as a stored signal, and
recorded 200 15-second spots from another TV broadcast as queries. Thus, the task was to
detect and locate specific commercial spots from 200 consecutive hours of TV recording.
Each spot occurred 2-30 times in the stored signal. Each signal was first digitized at a
32 kHz sampling frequency and 16 bit quantization accuracy. The bit rate for the MP3
encoding was 56 kbps. We extracted base features from each audio signal using a 7-channel
second-order IIR band-pass filter with $Q=10$. The center frequencies at the filter were
equally
spaced on a log frequency scale. The base features were calculated every 10 milliseconds from
a 60 millisecond window. The base feature vectors were quantized by using the VQ codebook
with 128 codewords, and histograms were created based on the scheme of the TAS method.
Therefore, the histogram dimension was 128. We implemented the feature sampling described
in Section \ref{subsec:piecewise:sampling} and the sampling duration was $a=50$.
The tests were carried out on a PC (Pentium 4 2.0 GHz).

\subsection{Search speed}

We first measured the CPU time and the number of matches in the search. The search time
we measured in this test comprised only the CPU time in the search stage shown in Section
\ref{sec:proposed}. This means that the search time did not include the CPU time for
any procedures in the preparation stage such as base feature extraction, histogram
creation, or histogram dimension reduction for the stored signal. The search threshold
was adjusted to $\theta=85$ so that there were no false detections or false dismissals.
We compared the following methods:
\renewcommand{\labelenumi}{(\roman{enumi})}
\begin{enumerate}
\item The TAS method (baseline).
\item The proposed search method without the projection distance being
      embedded in the compressed features.
\item The proposed search method.
\end{enumerate}

\begin{figure}[t]
  \begin{center}
    \includegraphics[width=8cm]{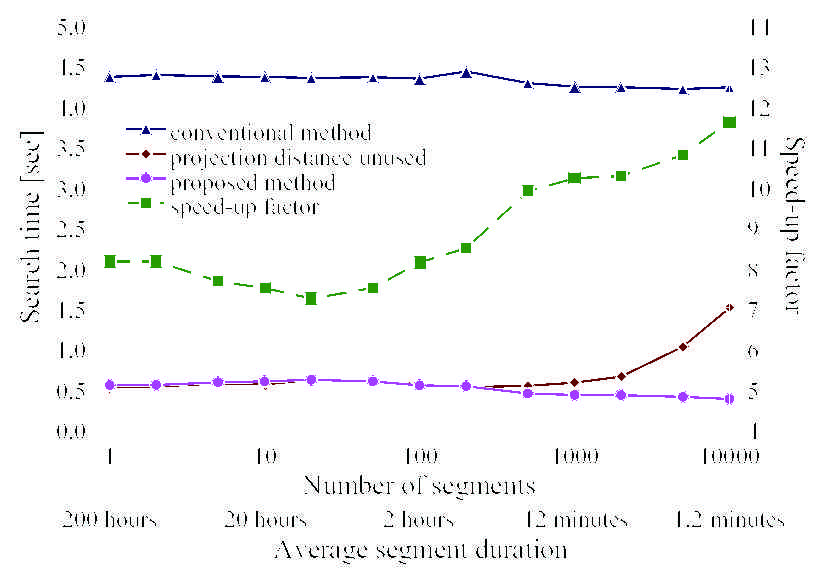}
  \end{center}
  \caption{Relationship between average segment duration and search speed measured by
           the CPU time in the search:
           (Horizontal axis) Average segment duration [200 hours - 1.2 minutes], which
           corresponds to the number of segments [1 - 10000],
           (Vertical axis) the CPU time in the search and the speed-up factor.}
  \label{graph:exp:piecewise:searchtime1}
\end{figure}
\begin{figure}[t]
  \begin{center}
    \includegraphics[width=8cm]{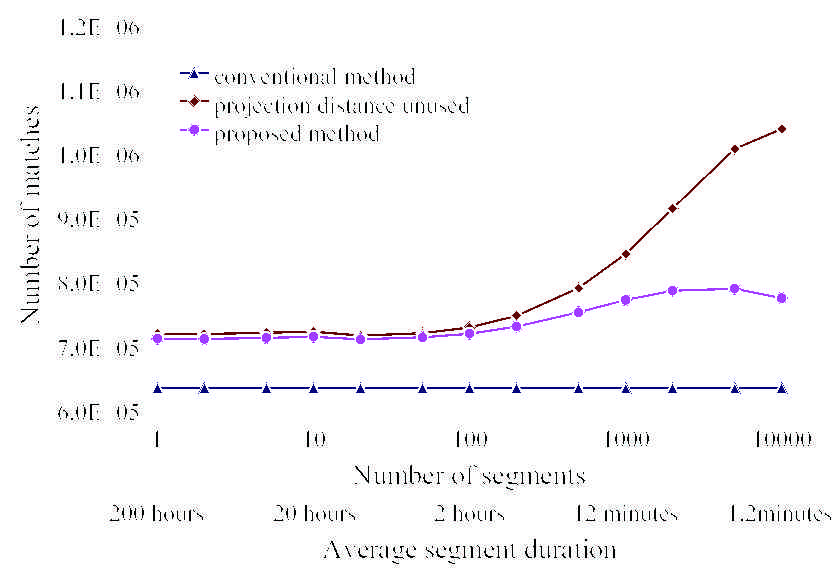}
  \end{center}
  \caption{Relationship between average segment duration and search speed measured by
           number of matches:
           (Horizontal axis) Average segment duration [200 hours - 1.2 minutes], which
           corresponds to the number of segments [1 - 10000],
           (Vertical axis) Number of matches.}
  \label{graph:exp:piecewise:searchcount1}
\end{figure}

We first examined the relationships between the average segment duration (equivalent to
the number of segments), the search time, and the number of matches. The following
parameters were set for feature-dimension reduction: The contribution threshold was
$\sigma=0.9$. The width of the shiftable range for dynamic segmentation was 500.

Fig. \ref{graph:exp:piecewise:searchtime1} shows the relationship between the average
segment duration and the search time, where the ratio of the search speed of the proposed
method to that of the TAS method (conventional method in the figure) is called the
{\it speed-up factor}. Also, Fig. \ref{graph:exp:piecewise:searchcount1} shows the
relationship between the average segment duration and the number of matches.
Although the proposed method only slightly increased the number of matches, it greatly
reduced the search time. This is because it greatly reduced the calculation cost per
match owing to
feature-dimension reduction. For example, the proposed method reduced the search time to
almost $1/12$ when the segment duration was 1.2 minutes (i.e. the number of segments was
$10000$). As mentioned in Section \ref{subsec:piecewise:sampling}, the feature sampling
technique also contributed to the acceleration of the search, and the effect is similar
to histogram skipping. Considering the dimension reduction performance results described
later, we found that those effects were greater than that caused by dimension
reduction for large segment durations (i.e. a small number of segments). This is examined
in detail in the next section. We also found that the proposed method reduced the
search time and the number of matches when the distance bounding technique was
incorporated, especially when there were a large number of segments.

\section{Discussion}
\label{sec:discuss}

The previous section described the experimental results solely in terms of search speed
and the advantages of the proposed method compared with the previous method. This section
provides further discussion of the advantages and shortcomings of the proposed method
as well as additional experimental results.

\begin{figure}[t]
  \begin{center}
    \includegraphics[width=8cm]{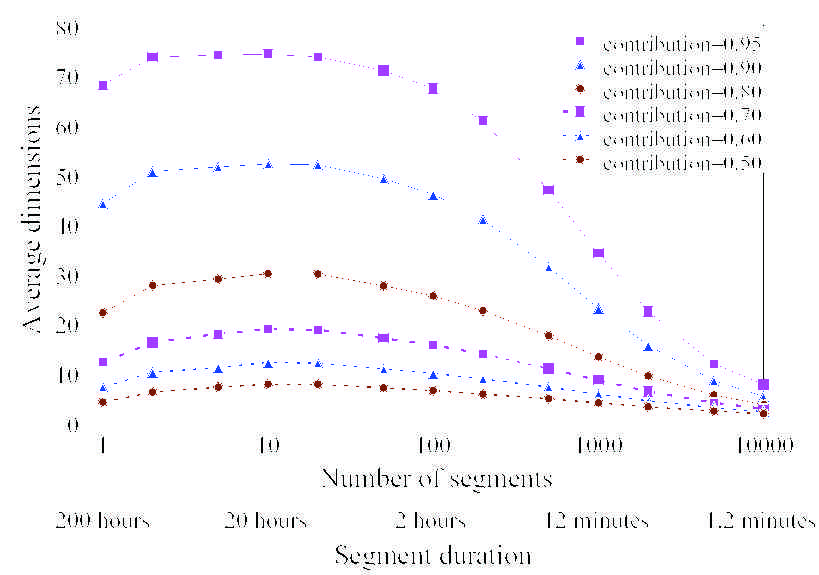}
  \end{center}
  \caption{Dimension reduction performance based on contribution rates:
           (Horizontal axis) Segment duration [200 hours - 1.2 minutes], which
           corresponds to the number of segments [1 - 10000]
           (Vertical axis) Average dimensionality of projected features per sample.}
  \label{graph:exp:compress1-1}
\end{figure}

We first deal with the dimension reduction performance derived from the segment-based KL
transform. We employed equi-partitioning to obtain segments, which means that we did not
incorporate the dynamic segmentation technique.
Fig. \ref{graph:exp:compress1-1} shows the experimental result. The proposed method
monotonically reduced the dimensions as the number of segments increased if the segment
duration was shorter than 10 hours (the number of segments $M\ge 20$). We can see that
the proposed method reduced the dimensions, for example, to $1/25$ of the original
histograms when the contribution threshold was $0.90$ and the segment duration was 1.2
minutes (the number of segments was $10000$). The average dimensions did not decrease as
the number of segments increased if the number of segments was relatively small. This is
because we decided the number of subspace bases based on the contribution rates.

\begin{figure}[t]
  \begin{center}
    \includegraphics[width=8cm]{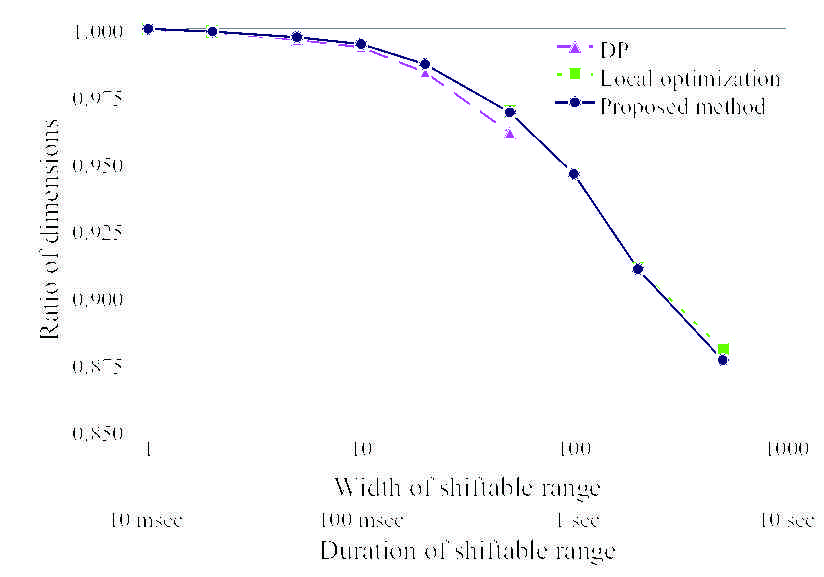}
  \end{center}
  \caption{Dimension reduction performance of dynamic segmentation
           [Number of segments$=1000$, contribution rate$=0.9$]:
           (Horizontal axis) Width of shiftable range [1 - 5000]
           (Vertical axis) Proportion of the dimensionality derived from dynamic
           segmentation compared with that obtained in the initial state
           (i.e. equi-partitioning).}
  \label{graph:exp:compress3-1}
\end{figure}

Next, we deal with the dimension reduction performance derived from the dynamic
segmentation technique. The initial positions of the segment boundaries were set by
equi-partitioning. The duration of segments obtained by equi-partitioning was 12 minutes
(i.e. there were $1000$ segments). Fig. \ref{graph:exp:compress3-1} shows the
result. The proposed method further reduced the feature dimensionality to $87.5\%$ of its
initial value, which is almost the same level of performance as when only the local
search was utilized. We were unable to calculate the average dimensionality when using DP
because of the substantial amount of calculation, as described later. When the shiftable
range was relatively narrow, the dynamic segmentation performance was almost the same as
that of DP.

Here, we review the search speed performance shown in Fig.
\ref{graph:exp:piecewise:searchtime1}. It should be noted that three
techniques in our proposed method contributed to speeding up the search, namely
feature-dimension reduction, distance bounding and feature sampling. When the number of
segments was relatively small, the speed-up factor was much larger than the ratio of the
dimension of the compressed features to that of the original histograms, which can be seen in
Figs. \ref{graph:exp:piecewise:searchtime1}, \ref{graph:exp:compress1-1} and
\ref{graph:exp:compress3-1}. This implies that the feature sampling technique dominated
the search performance in this case. On the
other hand, when the number of segments was relatively large, the proposed search method
did not greatly improve the search speed compared with the dimension reduction
performance. This implies that the feature sampling technique degraded the
search performance. In this case, the distance bounding technique mainly contributed to
the improvement of the search performance as seen in Fig.
\ref{graph:exp:piecewise:searchtime1}.

\begin{figure}[t]
  \begin{center}
    \includegraphics[width=8cm]{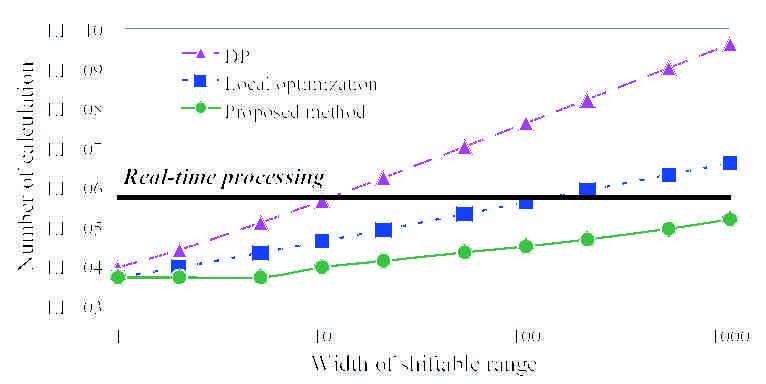}
  \end{center}
  \caption{
    Amount of computation for dynamic segmentation
    [Number of segments$=1000$, contribution rate$=0.9$]
    (Horizontal axis) Width of shiftable range [1 - 1000]
    (Vertical axis) Total number of PCA calculations needed to obtain the representation,
    where the horizontal line along with ``Real-time processing'' indicates that the
    computational time is almost the same as the duration of the stored signal.
  }
  \label{graph:exp:calcdynamic}
\end{figure}

Lastly, we discuss the amount of calculation necessary for dynamic segmentation. We again
note that although dynamic segmentation can be executed prior to providing a query
signal, the computational time must be at worst smaller than the duration of the stored
signal from the viewpoint of practical applicability. We adopted the total number of
dimension calculations needed to obtain the dimensions of the segments as a measure for
comparing the calculation cost in the same way as in Section \ref{sec:dynamic}. Fig.
\ref{graph:exp:calcdynamic} shows the estimated calculation cost for each dynamic
segmentation method. We compared our method incorporating local optimization and
coarse-to-fine detection with the DP-based method and a case where only the local
optimization technique was incorporated. The horizontal line along with ``Real-time
processing'' indicates that the computational time is almost the same as the duration of
the signal. The proposed method required much less computation than with DP or local
optimization. For example, when the width of the shiftable range was 500, the calculation
cost of the proposed method was $1/5000$ that of DP and $1/10$ that with local
optimization. We note that in this experiment, the calculation cost of the proposed
method is less than the duration of the stored signal, while those of the other two
methods are much longer.

\section{Concluding remarks}\label{sec:conclusion}

This paper proposed a method for undertaking quick similarity-based searches of an
audio signal to detect and locate similar segments to a given audio clip. The proposed
method was built on the TAS method, where audio segments are modeled by using histograms.
With the proposed method, the histograms are compressed based on a piecewise linear
representation of histogram sequences. We introduce dynamic segmentation, which divides
histogram sequences into segments of variable lengths. We also addressed the quick
suboptimal partitioning of the histogram sequences along with local optimization and
coarse-to-fine detection techniques. Experiments revealed significant improvements in
search speed. For example, the proposed method reduced the total search time to
approximately $1/12$, and detected the query in about $0.3$ seconds from a
200-hour audio database. Although this paper focused on audio signal retrieval, the
proposed method can be easily applied to video signal retrieval
\cite{piecewiselinear_icassp,dynamicsegmentation_icme}. Although the method proposed in
this paper is founded on the TAS method, we expect that some of the techniques we have
described could be used in conjunction with other similarity-based search methods (e.g.
\cite{QuickSearch,sugiyama,ModifiedTAS:Yuan,ModifiedTAS:Yuan2}) or a speech/music
discriminator \cite{BarbedoLopes}. Future work includes the implementation of indexing
methods suitable for piecewise linear representation, and the dynamic determination of
the initial segmentation, both of which have the potential to improve the search
performance further.

\appendices

\section{Proof of Theorem \ref{theorem:bound}}

First, let us define
\begin{eqnarray*}
  && \z_Q \eqdef p(\x_Q),\quad
     \z_S \eqdef p(\x_S),\\ 
  && \wh{\x}_Q \eqdef q(\z_Q) = q(p(\x_Q)),\quad
     \wh{\x}_S \eqdef q(\z_S) = q(p(\x_S)),\\
  && \delta_Q \eqdef \delta(p,\x_Q),\quad
     \delta_S \eqdef \delta(p,\x_S).
\end{eqnarray*}
We note that for any histogram $\x\in\cN^n$, $\wh{\x}=q(p(\x))$ is the projection of $\x$
into the subspace defined by the map $p(\cdot)$, and therefore $\x-\wh{\x}$ is a normal
vector of the subspace of $p(\cdot)$. Also, we note that $\|\x-\wh{\x}\|=\delta(p,\x)$
and $\wh{\x}$ is on the subspace of $p(\cdot)$. For two vectors $\x_1$ and $\x_2$, their
inner product is denoted as $\x_1\cdot\x_2$. Then, we obtain
\begin{eqnarray}
  \lefteqn{\|\x_Q-\x_S\|^2}\nonumber\\
  & = & \|(\x_Q-\wh{\x}_Q)-(\x_S-\wh{\x}_S)+(\wh{\x}_Q-\wh{\x}_S)\|^2\nonumber\\
  & = & \|\x_Q-\wh{\x}_Q\|^2+\|\x_S-\wh{\x}_S\|^2+\|\wh{\x}_Q-\wh{\x}_S\|^2\nonumber\\
  &   & -2(\x_Q-\wh{\x}_Q)\cdot(\x_S-\wh{\x}_S)
        +2(\x_Q-\wh{\x}_Q)\cdot(\wh{\x}_Q-\wh{\x}_S)\nonumber\\
  &   & -2(\x_S-\wh{\x}_S)\cdot(\wh{\x}_Q-\wh{\x}_S)\nonumber\\
  & = & \delta(p,\x_Q)^2+\delta(p,\x_S)^2+\|\wh{\x}_Q-\wh{\x}_S\|^2\nonumber\\
  &   & -2(\x_Q-\wh{\x}_Q)\cdot(\x_S-\wh{x}_S)\label{eq:proof:app:1}\\
  &\ge& \delta(p,\x_Q)^2+\delta(p,\x_S)^2+\|\wh{\x}_Q-\wh{\x}_S\|^2\nonumber\\
  &   & -2\delta(p,\x_Q)\cdot\delta(p,\x_S)\label{eq:proof:app:2}\\
  & = & \{\delta(p,\x_Q)-\delta(p,\x_S)\}^2+\|\z_Q-\z_S\|^2\nonumber\\
  & = & \|\y_Q-\y_S\|^2,\nonumber
\end{eqnarray}
where Eq. (\ref{eq:proof:app:1}) comes from the fact that any vector on a subspace and
the normal vector of the subspace are mutually orthogonal, and Eq.
(\ref{eq:proof:app:2}) from the definition of inner product.
This concludes the proof of Theorem \ref{theorem:bound}.

\section{Proof of Theorem \ref{theorem:error}}

The notations used in the previous section are also employed here. When the
projected features $\z_Q$, $\z_S$ and the projection distances
\begin{eqnarray*}
  \delta_Q\eqdef\delta(p,\x_Q),\quad \delta_S\eqdef\delta(p,\x_S)
\end{eqnarray*}
are given, we can obtain the distance between the original features as follows:
\begin{eqnarray}
  \lefteqn{\|\x_Q-\x_S\|^2}\nonumber\\
  &=& \|\z_Q-\z_S\|^2+\delta_Q^2+\delta_S^2\nonumber\\
  & & -(\x_Q-q(\z_Q))\cdot(\x_S-q(\z_S))\label{eq:proof:app:3}\\
  &=& \|\z_Q-\z_S\|^2+\delta_Q^2+\delta_S^2-2\delta_Q\delta_S\cos\phi,\nonumber
\end{eqnarray}
where Eq. (\ref{eq:proof:app:3}) is derived from Eq. (\ref{eq:proof:app:1}) and $\phi$ is the
angle between $\x_Q-q(\z_Q)$ and $\x_S-q(\z_S)$. From the assumption that random
variables $\X_S$ and $\X_Q$ corresponding to original histograms $\x_S$ and $\x_Q$ are
distributed independently and uniformly in the set $\cA$, the following equation
is obtained:
\begin{eqnarray}
  \lefteqn{E\left[\|\X_Q-\X_S\|^2-\|\z_Q-\z_S\|^2\right]}\nonumber\\
  & = & \int_0^\pi(\delta_Q^2+\delta_S^2-2\delta_Q\delta_S\cos\phi)\nonumber\\
  &   & \hspace{6mm}\frac{S_{n-m-1}(\delta_S\sin\phi)}{S_{n-m}(\delta_S)}
        |d(\delta_S\cos\phi)|, \label{eq:detail1}
\end{eqnarray}
where $S_k(R)$ represents the surface area of a $k$-dimensional hypersphere with radius
$R$, and can be calculated as follows:
\begin{eqnarray}
  S_k(R) &=& k\frac{\pi^{k/2}}{(k/2)!}R^{k-1} \label{eq:radius}
\end{eqnarray}
Substituting Eq. (\ref{eq:radius}) into Eq. (\ref{eq:detail1}), we obtain
\begin{eqnarray*}
  \lefteqn{E\left[\|\X_Q-\X_S\|^2-\|\z_Q-\z_S\|^2\right]}\nonumber\\
  &=&       \frac{n-m-1}{n-m}(\delta_Q^2+\delta_S^2)\nonumber\\
  &\approx& \frac{n-m-1}{n-m}\delta_Q^2,
\end{eqnarray*}
where the last approximation comes from the fact that $\delta_Q\gg\delta_D$.
Also, from Eq. (\ref{eq:distlb}) we have
\begin{eqnarray*}
  \|\x_Q-\x_S\|^2-\|\y_Q-\y_S\|^2 &=& 2\delta_Q\delta_S(1-\cos\phi).
\end{eqnarray*}
Therefore, we derive the following equation in the same way:
\begin{eqnarray*}
  \lefteqn{E\left[\|\X_Q-\X_S\|^2-\|\y_Q-\y_S\|^2\right]}\\
    & = & 2\frac{n-m-1}{n-m}\delta_Q\delta_S\\
    &\ll& E\left[\|\X_Q-\X_S\|^2-\|\z_Q-\z_S\|^2\right].
\end{eqnarray*}

\section*{Acknowledgments}
The authors are grateful to Prof. Yoshinao Shiraki of Shonan Institute of Technology for
valuable discussions and comments, which led to an improvement in the research. The
authors also thank Dr. Yoshinobu Tonomura, Dr. Hiromi Nakaiwa, Dr. Shoji Makino and Dr.
Junji Yamato of NTT Communication Science Laboratories for their support. Lastly, the
authors thank the associate editor Dr. Michael Goodwin and the anonymous reviewers for
their valuable comments.

\bibliographystyle{bib/IEEEtran} 
\bibliography{bib/IEEEabrv,bib/defs,bib/retrieval,bib/it}

\end{document}